\documentclass[10pt]{article}
\usepackage{graphicx,floatflt,amssymb,epsf}
\def\issue(#1,#2,#3){#1 (#3) #2} 
 
\def\opcit(#1){ {\em op. cit.}, #1} 
\def\etal {\em et al.} 

\def\APP(#1,#2,#3){Acta Phys.\ Polon.\ \issue(#1,#2,#3)} 
\def\ARNPS(#1,#2,#3){Ann.\ Rev.\ Nucl.\ Part.\ Sci.\ \issue(#1,#2,#3)} 
\def\CPC(#1,#2,#3){Comp.\ Phys.\ Comm.\ \issue(#1,#2,#3)} 
\def\CIP(#1,#2,#3){Comput.\ Phys.\ \issue(#1,#2,#3)} 
\def\EPJC(#1,#2,#3){Eur.\ Phys.\ J.\ C\ \issue(#1,#2,#3)} 
\def\EPJD(#1,#2,#3){Eur.\ Phys.\ J. Direct\ C\ \issue(#1,#2,#3)} 
\def\IEEETNS(#1,#2,#3){IEEE Trans.\ Nucl.\ Sci.\ \issue(#1,#2,#3)} 
\def\IJMP(#1,#2,#3){Int.\ J.\ Mod.\ Phys. \issue(#1,#2,#3)} 
\def\JHEP(#1,#2,#3){J.\ High Energy Phys. \issue(#1,#2,#3)} 
\def\MPL(#1,#2,#3){Mod.\ Phys.\ Lett.\ \issue(#1,#2,#3)} 
\def\NP(#1,#2,#3){Nucl.\ Phys.\ \issue(#1,#2,#3)} 
\def\NIM(#1,#2,#3){Nucl.\ Instrum.\ Meth.\ \issue(#1,#2,#3)} 
\def\PL(#1,#2,#3){Phys.\ Lett.\ \issue(#1,#2,#3)} 
\def\PRD(#1,#2,#3){Phys.\ Rev.\ D \issue(#1,#2,#3)} 
\def\PRL(#1,#2,#3){Phys.\ Rev.\ Lett.\ \issue(#1,#2,#3)} 
\def\SJNP(#1,#2,#3){Sov.\ J. Nucl.\ Phys.\ \issue(#1,#2,#3)} 
\def\ZPC(#1,#2,#3){Zeit.\ Phys.\ C \issue(#1,#2,#3)} 
 
\def\bra {\langle} 
\def\ket {\rangle} 
 
\def\l {\lambda}

\def\r {\rightarrow}

\def\bar {\overline} 
\def\bbbar {B^0-\overline{B^0}}

\def\be {\begin{equation}} 
\def\ee {\end{equation}} 
\def\bea {\begin{eqnarray}} 
\def\eea {\end{eqnarray}} 
\def\n {\nonumber} 
\def\bc {\begin{center}} 
\def\ec {\end{center}}

\begin{document}
\begin{flushright}
\texttt{hep-ph/0212059}\\
SINP/TNP/02-32\\
\end{flushright}

\vskip 50pt

\begin{center}
{\Large \bf \boldmath $R$-Parity violation in 
\boldmath $B\r\pi^+\pi^-$ decay}
\\
\vspace*{1cm}
\renewcommand{\thefootnote}{\fnsymbol{footnote}}
{\large {\sf Gautam Bhattacharyya ${}^1$}, {\sf Amitava Datta ${}^2$},
and {\sf Anirban Kundu ${}^3$}
} \\
\vspace{10pt}
{\small 1. Saha Institute of Nuclear Physics, 1/AF Bidhan 
        Nagar, Kolkata 700064, India \\ 
   2. Department of Physics, Jadavpur University, Kolkata 700032,
      India \\
   3. Department of Physics, University of Calcutta, 92 A.P.C.
        Road, Kolkata 700009, India  
} 

\normalsize
\end{center}

\begin{abstract}

We consider the impact of $R$-parity violating suprsymmetry in the
nonleptonic decay $B\to\pi^+\pi^-$. This is one of the rare instances
where new physics contributes to both $\bbbar$ mixing and $B\to\pi\pi$
decay. We do a numerical analysis to capture the interplay between
these two effects and place constraints on the relevant parameter
space.

\vskip 5pt \noindent
\texttt{PACS Nos:~11.30.Fs, 12.60.Jv, 13.25.Hw} \\
\texttt{Key Words:~~$R$-parity violation, $B$ decays}
\end{abstract}

\renewcommand{\thesection}{\Roman{section}}
\setcounter{footnote}{0}
\renewcommand{\thefootnote}{\arabic{footnote}}

\section{Introduction} 
In the standard model (SM), baryon and lepton numbers ($B$ and $L$,
respectively) are automatically conserved, while in supersymmetry (SUSY)
neither gauge symmetry nor any such fundamental principles tells us as to why
these discrete symmetries should be exactly respected. This prompts us to
define in supersymmetric theories a discrete quantum number called $R$-parity
as $R = (-1)^{(3B+L+2S)}$ \cite{def_rpar,intro_rpar}, where $S$ is the spin of
the particle.  In fact, it is not fair to be totally dictated by theoretical
prejudice and {\em ab initio} abandon the $R$-parity violating (RPV) terms. A
rather open-minded approach would be to keep the RPV interactions in the
theory and constrain them from observations/ non-observations at different
experiments. There is a rich phenomenological consequence of $R$-parity
violation. When $R$-parity is violated, the lightest supersymmetric particle
is no longer stable and superparticles need not be produced in pairs. As a
result, one needs to reformulate the SUSY search strategies. A complete set of
RPV interactions introduces 48 new parameters into the theory, and so it is
important to constrain them from existing data. Upper bounds on RPV couplings
emerge from proton stability, $n$--$\bar{n}$ oscillation, neutrino masses and
mixings, charged current universality, atomic parity violation, $Z$ pole
observables, meson decays and mixings, etc., a thorough account of which has
been reviewed in \cite{rpv}.

During the last few years many new data on branching ratios (BRs) and CP
asymmetries in different $B$ decay channels are coming from the
$B$ factories. It is therefore worthwhile to examine the consequences of these
data for RPV models. Let us first discuss why $B$ decays in RPV scenario are
interesting. In the $R$-parity conserving limit, leading SUSY contributions to
$B$ decays would come from penguins and/or boxes.  As has been noted by
several authors, e.g. \cite{barbieri}, such SUSY loops are either suppressed,
or at best comparable, with respect to the SM penguins on account of the
heaviness of SUSY spectrum.  {\em On the contrary, RPV might trigger such
decays at tree level with suitable couplings turned on, and there lies its
importance}. Moreover, such RPV couplings are complex, and their phases play
their part in inducing new contribution to CP violation.  As an example,
processes which are purely penguin driven in the SM (e.g. $b \to sd\bar{d}$ or
$b \to ss\bar{s}$), may not receive any appreciable corrections from SUSY
loops, but tree level RPV contributions to them might be significant. Even
otherwise, in processes like $b \to c\bar{c}s$ which is dominated by the SM
tree diagram, a RPV tree contribution might have an effect in the extraction
of the angle $\beta$ of the unitarity triangle. All these effects in the
context of $B$ decaying into $\pi K$, $\phi K$, $J/\psi K_S$ final states have
been studied by many authors \cite{rpvcp,alak}.

{\em In this paper we shall restrict ourselves to $B \to \pi^+ \pi^-$ decay,
and comment on other $\pi\pi$ modes ($\pi^0\pi^0$, $\pi^0\pi^\pm$).}  The
dominant SM contribution to $B \to \pi^+ \pi^-$ decay comes from a tree graph
which has a suppression from $V_{ub}$.  The RPV amplitude is driven by the
product $\lambda'_{i11}\lambda'^*_{i13}$.  For $i=$ 2 and 3, the existing
bound on this combination is rather modest ($|\lambda'_{i11} \lambda'_{i13}|<
3.6 \times 10^{-3}$) \cite{gg-arc}.  The propagator of the RPV tree diagram is
a slepton whose mass can be in the 100 GeV range. Thus the RPV amplitude can
be comparable in size with the SM amplitude.  It may be recalled at this stage
that the presence of two interfering amplitudes of comparable magnitude is
essential for a large direct CP violating asymmetry.

Additionally, the above product coupling contributes to the $\bbbar$ mixing
through box graphs \cite{gg-arc}, and that makes the situation more tricky. In
fact, this is one of the very few situations where a new physics operator can
contribute to {\em both} mixing and decay with amplitudes comparable, or
possibly even larger, than the SM contributions. The angle $\beta$ of the
unitarity triangle is determined from the $\bbbar$ mixing phase $\phi_M$.  The
SM box diagram yields $\phi_M$ to be 2$\beta$.  Now that the
$\lambda'_{i11}\lambda'^*_{i13}$ combination is turned on, there is new
contribution {\em not only to decay into $\pi \pi$ final states but also to
$\bbbar$ mixing at the same time}.  {\em The twin r\^{o}le of these specific
product couplings for the $\pi \pi$ final states thus adds a new twist and
brings in significant calculational intricacies}.

Now once the mixing phase gets contaminated, the immediate concern is what
happens then to $B$ decays into, for example, $J/\psi K_S$ and $\phi K_S$
final states?  Supposing that only the above couplings, namely the
$\lambda'_{i11}\lambda'^*_{i13}$ combinations, are operative, there will not
be any new contribution to the decay diagrams for the above final states, but
the corresponding mixing induced asymmetries will get modified through new
contribution to $\phi_M$. The CP asymmetries in the above channels are
proportional to sin 2$\beta$ in the SM. Now both the asymmetries will measure
not the angle $\beta$ (defined as argument of $V_{td}$) but some other
$\beta_{eff}$ carrying traces of new physics.

Now, as mentioned before, the $\pi\pi$ final states from $B$ decays require
special attention as the given choice of RPV product couplings inducts new
effects {\em simultaneously into mixing and decay}.  Using the available data
set on BRs and CP asymmetries (to be reviewed in the next section), we obtain
useful constaints on the RPV product couplings. This is true even considering
the fact that the present experimental errors are still quite large. For
example, {\em we obtain a new bound $|\lambda'_{i11}\lambda'^*_{i13}| \leq
0.0022$ for a slepton mass of around 100 GeV, which is a marginal improvement
over its existing bound of 0.0036}.  However, our bound is more general since
we take into account the possibility of destructive interference between the
SM and the RPV box amplitudes where as the previous bound was derived by
saturating the experimental number by RPV box only. In addition, interesting
correlations among $\gamma$ (another angle of the unitarity triangle), the
weak phase of the RPV product coupling, and the strong phase difference
between the SM and RPV amplitudes emerge.  It is to be noted that our {\em
primary aim} is to put an upper bound on new physics parameters (RPV
couplings) using the standard technique of accomodating as much of new physics
as possible in the space between experimental data and SM predictions in the
$B\to\pi^+\pi^-$ channel.  In the process, we observe a simultaneous
enhancement of the BR in $\pi^0\pi^0$ final state which is a big bonus in view
of recent data showing large BR in this channel for which the SM prediction is
too low.

\section{Review of data and the relevant formulae}
Data is available on all $\pi\pi$ modes.  Let us first look at the BRs
(multiplied by $10^6$) as they stand in ICHEP 2004 \cite{hfag}:
\begin{eqnarray}
Br(B\to\pi^+\pi^-) &=& 4.6\pm 0.4;\nonumber\\
Br(B\to\pi^0\pi^0) &=& 1.51\pm 0.28;\\
Br(B^\pm\to\pi^\pm\pi^0) &=& 5.5\pm 0.6.\nonumber
\end{eqnarray}
We use $B$ to indicate a flavor-untagged $B^0$ or $\bar{B^0}$. Defining
\begin{eqnarray}
\l = \exp(-i2\beta)\bra \pi^+\pi^-|{\cal H}|\bar{B^0}\ket /
\bra \pi^+\pi^-|{\cal H}|B^0\ket, 
\end{eqnarray}
and the direct and mixing induced CP asymmetries as 
\begin{eqnarray}
a^d_{CP}=(1-|\l|^2)/(1+|\l|^2),~~~
a^m_{CP}=2{\rm Im}\l/(1+|\l|^2), 
\end{eqnarray}
we write their present experimental numbers \cite{hfag} as 
\begin{eqnarray}
S_{\pi\pi}&=&-a^m_{CP}=-0.74\pm 0.16,\nonumber\\
C_{\pi\pi}&=&a^d_{CP}=-0.46\pm 0.13,\nonumber\\
A_{CP}(\pi^+\pi^0) &=& 0.01\pm 0.07, \\
A_{CP}(\pi^0\pi^0) &=& -0.28\pm 0.39. \nonumber
\end{eqnarray}
The quantities $a^d_{CP}$ and $a^m_{CP}$ are obtained from the
time-dependent measurement on $B$: 
\bea 
a^{dm}_{\pi^+\pi^-}(t) &=&
{\Gamma(B^0(t)\r \pi^+\pi^-) - \Gamma(\bar{B^0}(t) \r \pi^+\pi^-)
\over \Gamma(B^0(t)\r \pi^+\pi^-) + \Gamma(\bar{B^0}(t) \r
\pi^+\pi^-)}\n\\ &=&a_{CP}^d\cos\Delta mt+a_{CP}^m\sin\Delta mt, 
\eea
where $\Delta m$ is the mass difference between the two mass
eigenstates.

To motivate our further discussions, let us assume  
temporarily that only two interfering 
amplitudes contribute to the $\bar{B^0}\r\pi^+\pi^-$ and denote them by 
$$ 
a_1\exp(i\phi_1)\exp(i\delta_1) \ \ \ {\rm and} \ \ \  
a_2\exp(i\phi_2)\exp(i\delta_2), 
$$ where $\phi_i$'s and $\delta_i$'s ($i=1,2$) are the weak and the strong
phases, respectively.  We also use the notation \be \Delta\delta =
\delta_2-\delta_1; \ \ \ \Delta\phi = \phi_2-\phi_1.  \ee We, however, wish to
emphasize that for the actual numerical analyses, we will take into account
not only the SM tree diagram but also the SM penguin amplitudes as
well\footnote{The SM amplitudes in the naive factorization approach are given
in Eq.~(A1) of Appendix-A in Ref.~\cite{ali}.}, in addition to new physics
contributions.
 
The observables $a_{CP}^{d}$ and $a^m_{CP}$ can be expressed in terms 
of the above parameters. One obtains 
\bea 
\label{acpd}
a_{CP}^{d} & = &{1-|\lambda_{\pi\pi}|^2\over 1+|\lambda_{\pi\pi}|^2} 
= {2a_1a_2\sin\Delta\phi\sin\Delta\delta\over a_1^2 + a_2^2 + 2a_1a_2 
\cos\Delta\phi\cos\Delta\delta}, \\ 
{\rm and} ~~~~
\label{acpm} 
a^m_{CP} &=& {2\ {\rm Im}\lambda_{\pi\pi}\over 
1+|\lambda_{\pi\pi}|^2}, 
~~~{\rm where}~~~\\ 
\lambda_{\pi\pi} &=&  
e^{-i\phi_M}{\bra \pi^+\pi^-|{\cal H}_{eff}|
\bar{B_q}\ket\over \bra \pi^+\pi^-|{\cal H}_{eff}|
B_q\ket} 
= \eta_{CP} e^{i(-\phi_M+2\phi_1)} 
{\left 
(a_1^2+a_2^2e^{2i\Delta\phi} +  
2a_1a_2e^{i\Delta\phi}\cos\Delta \delta\right) 
\over 
\left (a_1^2+a_2^2+2a_1a_2\cos(\Delta\delta-\Delta\phi)\right) }. 
\eea 
Here $\phi_M$ is the phase of the $B^0 - \bar{B^0}$ mixing amplitude (this may
include phases from the CKM elements as well as phases from new physics), and
$\eta_{CP}$ is the CP eigenvalue ($+1$) for the final state $\pi^+\pi^-$.
 
For the sake of completeness we also include the expressions for the 
BR$(B\r\pi^+\pi^-)$:  
\bea {\rm BR}(B^0 \r \pi^+ \pi^-) &\sim& a_1^2 + a_2^2 
+ 2a_1a_2\cos(\Delta\delta -\Delta\phi), \nonumber\\  
{\rm BR} (\overline{B^0} \r 
\pi^+ \pi^- ) &\sim& a_1^2 + a_2^2 + 2a_1a_2 \cos(\Delta\delta+ 
\Delta\phi),  
\eea  
where the phase space factors have been suppressed. When one averages over
these two terms, one obtains the expression in the denominator of
Eq.~(\ref{acpd}).

In the SM $a_1$ and $a_2$ are identified with the tree and the top-mediated
strong penguin amplitudes, respectively, so that $\phi_1=-\gamma$, $\phi_2 =
\beta$ and $\phi_M = 2\beta$.  One expects $a_2$ to be suppressed with respect
to $a_1$ due to the standard loop suppression factors.  This is an approximate
representation of the standard form of the amplitudes found in the literature
\cite{bfrs}, where both charm- and up-quark mediated strong penguins, as well
as the electroweak penguins, are taken into account. One usually decomposes
the top-penguin into terms proportional to $V_{ub}V^\ast_{ud}$ and
$V_{cb}V^\ast_{cd}$ using the unitarity relationship. The term proportional to
$V_{ub}V^\ast_{ud}$ is dumped with the tree amplitude, since all of them carry
the same weak phase, and this combination is usually called $T_c$. The terms
proportional to $V_{cb}V^\ast_{cd}$ does not have any weak phase in the
leading order of the Wolfenstein representation; this group is called $P_c$.
The electroweak penguins are expected to be much smaller than both $T_c$ and
$P_c$, and they can be neglected, or treated separately, depending upon the
precision required. In the approximation that the penguins are much smaller
than the tree amplitude, and among the penguins, the top-quark mediated one is
the dominant, we can use the numbers quoted in the literature for $T_c$ and
$P_c$ for $a_1$ and $a_2$, respectively.  As a standard prediction in all
theoretical models, one finds $|P_c/T_c|\sim 0.25-0.35$ \cite{amprefs}.  Thus
the observable $a_{CP}^{d}$ appears to be small in the SM.  Moreover, as we
shall see below, the measured BR for $\pi^+\pi^-$ mode turns out to be a bit
smaller than the SM prediction, the degree of discrepancy depending upon the
method of calculation.

Thus, the explanation of the BR($B \r \pi^+\pi^-$) data needs a destructive
interference between the two amplitudes leading to large
$|\cos\Delta\delta|$. On the other hand, large asymmetry as experimentally
measured requires large $|\sin\Delta\delta|$.  These twin requirements
constrain the parameter space.

\section{The \boldmath ${\bbbar}$ mixing and \boldmath ${B\r\pi^+\pi^-}$ decay 
in \boldmath ${R}$-parity violating supersymmetry}
It is well known that in order to avoid rapid proton decay one cannot have
both lepton number and baryon number violating RPV model, and we shall work
with a lepton number violating one. This leads to slepton/sneutrino
mediated$B$decays, and slepton/sneutrino/squark mediated $\bbbar$ mixing. We
start with the superpotential
\be
\label{w} 
{\cal W}_{\lambda'} = \lambda'_{ijk} L_i Q_j D^c_k, 
\ee 
where $i, j, k = 1, 2, 3$ are quark and lepton generation indices; $L$ and $Q$
are the SU(2)-doublet lepton and quark superfields and $D^c$ is the
SU(2)-singlet down-type quark superfield, respectively. For the process
$B\rightarrow \pi^+\pi^-$, the relevant four-Fermi operator is of the form \be
{\cal H}_{\lambda'} =
\frac{\lambda'_{i11}\lambda'^*_{i13}}{2m^2_{\tilde{e}_{L_i}}}
(\bar{u}\gamma^\mu P_L u)_8 (\bar{d}\gamma_\mu P_R b)_8 + {\rm h.c.}  \ee
where $P_R(P_L)= (1+(-)\gamma_5)/2$, and the subscript 8 denotes a color octet
combination. In the above formula $i$ is the generation index of the slepton.
The current bound on $\lambda'_{111}$ is too restrictive ($|\lambda'_{111}| <
3.5\times 10^{-4}$ \cite{rpv}), which rules out the possibility that this
coupling plays any significant role in$B$decays.  For $i=2$ or 3, the bound on
the product $\lambda'_{i11}\lambda'^*_{i13}$ is rather modest
($|\lambda'_{i11} \lambda'^*_{i13}|< 3.6 \times 10^{-3}$) \cite{gg-arc}.
Following the standard practice we shall assume that the RPV couplings are
hierarchical {\em i.e.}, only one combination of the couplings is numerically
significant. For simplicity we choose to ignore the RPV penguin contributions,
which are expected to be small even compared to the SM penguin amplitudes;
this follows from the smallness of the relevant RPV couplings compared to the
SM gauge couplings.  The bounds on the RPV couplings that we eventually derive
are insensitive to the inclusion of RPV penguins.

There is a much stronger bound on the product couplings of the type
$\lambda'_{i13}{\lambda'}_{i31} \leq 8.~10^{-8}$ from tree-level $\bbbar$
mixing (see \cite{rpv}); however, we consider only one type of product
coupling to be nonzero (namely, $\lambda'_{i11}\lambda'^*_{i13}$) in our
analysis.

Our discussion of $\bbbar$ mixing in the framework of an $L$-violating
RPV model follows that of \cite{ak-jps:3}. 
The off-diagonal element in the $2\times 2$ effective Hamiltonian
causes the $\bbbar$ mixing.  The mass difference
between the two mass eigenstates $\Delta M$ is given by (following
the convention of \cite{buras-fleischer})
\begin{equation}
\Delta M = 2|M_{12}|,
\end{equation}
with the valid approximation $|M_{12}|\gg |\Gamma_{12}|$.
Let the SM amplitude be
\begin{equation}
|M_{12}^{SM}|\exp(-2i\theta_{SM})
\end{equation}
where $\theta_{SM}= \beta $ for the ${\bbbar}$ system.  We follow the
$(\alpha, \beta,\gamma)$ convention for the unitarity triangle
\cite{buras-fleischer}.

If we have $n$ number of new physics (NP) amplitudes with weak 
phases $\theta_n$, one can
write
\begin{equation}
M_{12} = |M_{12}^{SM}|\exp(-2i\theta_{SM}) + \sum_{i=1}^n
|M_{12}^i|\exp(-2i\theta_i).
\end{equation}
This immediately gives the effective mixing phase $\theta_{eff}$ as
\begin{equation}
\theta_{eff} = {1\over 2}\arctan {|M_{12}^{SM}|\sin(2\theta_{SM}) +
\sum_i|M_{12}^i|\sin(2\theta_i)
\over |M_{12}^{SM}|\cos(2\theta_{SM}) + \sum_i|M_{12}^i|\cos(2\theta_i)},
\end{equation}
and the mass difference between mass eigenstates as
\begin{equation}
\Delta M  =  2\left[ |M_{12}^{SM}|^2 + \sum_i|M_{12}^i|^2
 + 2|M_{12}^{SM}|\sum_i |M_{12}^i|\cos 2(\theta_{SM}-\theta_i)
 +  2 \sum_i \sum_{j>i} |M_{12}^j||M_{12}^i|\cos 2(\theta_j-\theta_i)
\right]^{1/2}.
\end{equation}
These are going to be our basic formulae. The only task is to find 
$M_{12}^i$ and $\theta_i$. 

The SM mixing amplitude, dominated by the top-quark loop, is
\begin{eqnarray}
M_{12}^{SM}&\equiv& {\bra \bar{B^0}|H^{SM}_{eff}|B^0\ket\over
2m_B}\nonumber\\ &=& {G_F^2\over 6 \pi^2}(V_{td}V_{tb}^*)^2 \eta_B m_B
f_B^2 B_B m_W^2 S_0(x_t).  \label{b-sm}
\end{eqnarray}
where 
\begin{equation}
S_0(x) = {4x-11x^2+x^3\over 4(1-x)^2} - {3x^3\ln x \over 2(1-x)^3}.
\end{equation}

It is easy to check from the RPV superpotential that there are two different
kind of boxes that contribute to $\bbbar$ mixing: first, the one where one has
two sfermions flowing inside the loop, alongwith two SM fermions
\cite{decarlos-white}, and secondly, the one where one slepton, one $W$ (or
charged Higgs or Goldstone) and two up quarks complete the loop
\cite{gg-arc}. It is obvious that the first amplitude is proportional to the
product of four $\lambda'$ type couplings, and the second to the product of
two $\lambda'$ type couplings times $G_F$. We call them L4 and L2 boxes,
respectively. The detailed calculations including the QCD corrections at the
next-to-leading order (NLO), which we follow in our present analysis, may be
found in \cite{ak-jps:3}.

In the presence of RPV, the
$\Delta B = 2$ effective Hamiltonian can be written as
\begin{equation}
{\cal H}_{eff}^{\Delta B = 2} = \sum_{i=1}^5 c_i(\mu) O_i(\mu) +
\sum_{i=1}^3 \tilde c_i(\mu) \tilde O_i(\mu) + {\rm h.c.}
\end{equation}
where $\mu$ is the regularization scale, and
\begin{eqnarray}
O_1&=&(\bar b \gamma^\mu P_L d)_1(\bar b \gamma_\mu P_L d)_1,\nonumber\\
O_2&=&(\bar b P_R d)_1(\bar b  P_R d)_1,\nonumber\\
O_3&=&(\bar b P_R d)_8(\bar b  P_R d)_8, \\
O_4&=&(\bar b P_L d)_1(\bar b  P_R d)_1,\nonumber\\
O_5&=&(\bar b P_L d)_8(\bar b  P_R d)_8, \nonumber
   \label{operators}
\end{eqnarray}
where $P_{R(L)}=(1+(-)\gamma_5)/2$. The subscripts 1 and 8 indicate whether
the currents are in colour-singlet or in colour-octet combination. The $\tilde
O_i$s are obtained from corresponding $O_i$s by replacing $L\leftrightarrow
R$.  The Wilson coefficients $c_i$ and $\tilde{c_i}$ at $q^2=m_W^2$ and at the
low energy scale have been taken from \cite{becirevic}.

Now we come to the decay amplitude.  The matrix element of the RPV operator
for $B\r\pi^+\pi^-$ is given, using conventional factorization \cite{ali}, by
\be \bra \pi^+\pi^-|{\cal
H}_{\lambda'}|\overline{B^0}\ket = -{1\over 4}
\frac{\lambda'_{i11}\lambda'^*_{i13}}{m^2_{\tilde{e}_{L_i}}}
{m_\pi^2\over (m_d+m_u)(m_b-m_u)} f_\pi F_0^{B\r\pi}(m_\pi^2)
(m_B^2-m_\pi^2) 
\ee 
where $f_\pi$ is the pion decay constant and $F_0$ is the Bauer-Stech-Wirbel
(BSW) form factor.  We use conventional factorization \cite{ali} for
simplicity.

\section {Numerical results} 
We now turn to the numerical results. The calculational intricacies leading to
our results are two-fold.  First, we must bring in the RPV box diagram for
reasons discussed above.  The dependence on the weak phase associated with the
complex RPV product couplings $\lambda'_{i11} {\lambda'^*_{i13}}$ is therefore
rather involved and intriguing. Second, although for illustration we have
often discussed in the text the interference of SM tree and RPV tree
amplitudes only, leading to CP asymmetries in the $B \to \pi \pi$ channel, for
actual numerical details we have included the SM penguin as the third
interfering amplitude, and its effect is numerically non-negligible.  Although
the introduction of RPV introduces a few more parameters (see below), we now
have sufficient number of constraints. As a consequence, the new results, in
particular the constraints on the magnitude of $\sin \Delta \delta$ and the
product coupling $\lambda'_{i11}\lambda'^*_{i13}$ have a very restrictive
pattern.

The RPV model introduces four extra parameters compared to the SM: (a) the
left slepton and sneutrino masses, which are equal up to the SU(2) breaking
D-terms, (b) the magnitude of the product $\lambda'_{i11}\lambda'^*_{i13}$
(which according to our convention can be either positive or negative), (c)
the phase of this product, hereafter called $\phi$ or the weak RPV phase,
which can have any value between 0 and $\pi$ to maintain consistency with the
sign convention in (b), and (d) the strong phase between the SM tree and the
RPV amplitudes varying between 0 and $2\pi$. We fix the slepton mass at 100
GeV which is consistent with the current bound coming from direct
searches. The squarks are taken to be degenerate with the sleptons. Even
though that is unrealistic from the Tevatron data, the numbers change only
marginally. The magnitude of the product coupling and the remaining two
parameters are randomly varied within their allowed ranges or bounds.
 
In order to carry out the numerical analysis we need some more inputs like
quark masses, form factors and the relevant CKM elements.  We use $m_u = 4.2$
MeV, $m_d = 7.6$ MeV, $m_b = 4.88$ GeV, pion decay constant $f_\pi = 132$ MeV,
and the decay form factor in the BSW model $F_0^{B\r\pi} (m_\pi^2) = 0.39$.
The relevant Wilson coefficients for $b\to d$ transition in the LO scheme are
taken from \cite{ali}.  We expect the strong phase difference between the SM
tree and the SM penguin amplitudes to be small, but do not impose it as a
constraint.  The CKM parameters whose values are not precisely known have been
varied randomly within the range allowed by the CKM fit \cite{ckmfitter2004}.
In particular $V_{td}$ is allowed to lie in the range between 0.0030 and
0.0096. Arguably such ranges may change in the presence of RPV, since the
$\bbbar$ mixing amplitude and the resulting mass difference of $B$ meson mass
eigenstates ($\Delta m_{B^0}$), an important ingredient of the CKM fit, are
affected for reasons discussed above. In order to compensate for the
restricted inputs we have not constrained the weak phase $\gamma$ within the
SM range, but varied it randomly in the entire range of 0 to $\pi$
\footnote{The angle $\gamma$ is expected to lie in the first quadrant from the
CKM fit when $\bbbar$ mixing data, {\em i.e.}, $\Delta m_{B^0}$, is taken into
account. Once we entertain the possibility of new physics in $\bbbar$ mixing,
that constraint is no longer applicable, and there is a second possible
solution, with $\gamma$ in the second quadrant \cite{ckmfitter2004}.}.  The
other important input parameter $\sin (2\beta)$ has been varied in the range
allowed by the CKM fit: $0.725 \pm 0.033$ at 68\% C.L.
\cite{ckmfitter2004}. The justification for using such limits is that the
usual CKM fit without the direct measurement of CP asymmetries yields $\sin
(2\beta)$ very close to this range.  We have checked that none of our results,
apart from the allowed range of RPV weak phase, depends sensitively on the
choice of the angle $\beta$, and thus this analysis holds for some other
slightly different CKM fits too.

The origin of the strong phase can at present only be modeled. In naive
factorization, admittedly, there cannot be any strong phase between two tree
amplitudes.  But in pQCD, dynamical enhancement of annihilation and exchange
topologies play a significant part in generating a significant strong phase
difference.  Our approach has been the following: (i) we have used
factorization model as a simplistic approach to calculate amplitudes, {\em
but} (ii) we have treated the strong phase difference between the dominant
interfering amplitudes as a {\em phenomenological parameter}.

\begin{figure}[htbp] 
\vspace{-10pt}
\centerline{\hspace{-3.3mm} 
\rotatebox{-90}{\epsfxsize=8cm\epsfbox{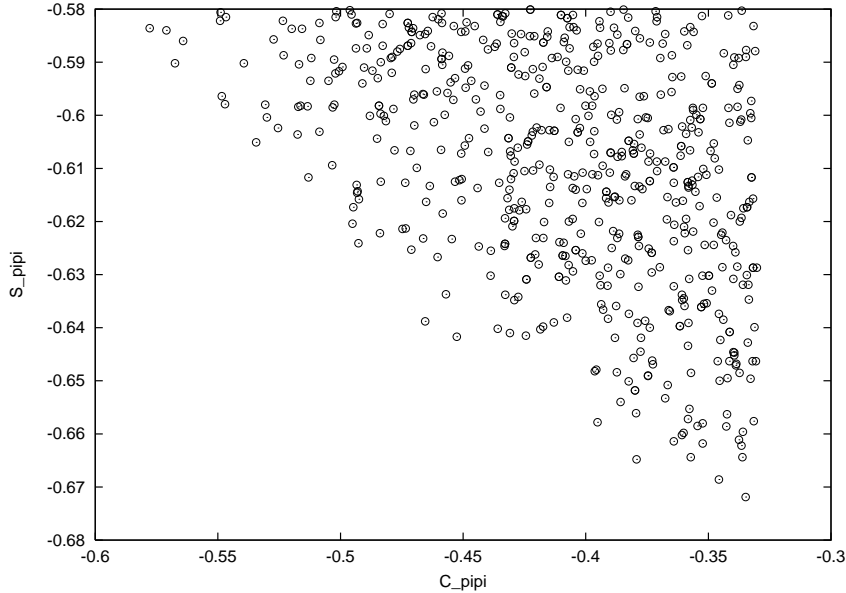}}} 
\hspace{3.3cm}\caption[]{\small Correlation of $S_{\pi\pi}$ and
$C_{\pi\pi}$.}  \protect\label{fig1} 
\end{figure} 
 
As stated above the effective Hamiltonian in Eq.\ (13) also leads to a pair of
new box amplitudes for $\bbbar$ mixing, {\em viz.}, L2 and L4.  The first kind
has two $\lambda'$ vertices, two SU(2) gauge couplings, and two up quarks, one
slepton and one $W$ inside the box. We have neglected, for simplicity, the
imaginary part which arises in this diagram from cutting the up quark lines.
The second type has four $\lambda'$ vertices.  Neglecting the SM box
completely, and taking the product coupling to be real, the authors in
\cite{gg-arc} found the conservative bound $|\lambda'_{i11}\lambda'_{i13}|
\leq 3.6\times 10^{-3}$. We, on the other hand, take into account the SM box
and the possible phase of the RPV product coupling which is randomly varied
over the range already given.  This, as discussed above, modifies the phase
$\phi_M$ from its SM value of $2\beta$ to $2\beta_{eff}$.  We now impose the
constraint that $\sin(2\beta_{eff})$ should satisfy the observed CP-asymmetry
in the $B\rightarrow J/\psi K_S$ channel ({\em i.e.}, $\beta_{eff}$, which is
a combination of $\beta$, RPV weak phase $\phi$, and the box amplitudes,
should satisfy $0.692 \leq \sin(2\beta_{eff})\leq 0.758$).
 
We next list all the constraints imposed in our study of the allowed space of
the RPV parameters: (i) $\Delta m_{B^0}$, (ii) CP asymmetry from the decay
$B\r J/\psi K_S$, (iii) BR($B\r\pi^+\pi^-$), and (iv) the asymmetries
$C_{\pi\pi}$ and $S_{\pi\pi}$.  In addition, we also impose the model
independent constraint $S_{\pi\pi}^2+C_{\pi\pi}^2<1$.  All the experimental
numbers are taken at 1$\sigma$.
 
The random variation of the parameters subject to the constraints as
discussed above leads to the scatter plots displayed in Figures 1 and
2. The following salient features are to be noted.

1. From Figure 1, one can see that though $C_{\pi\pi}$ can be accomodated over
its entire range, $S_{\pi\pi}$ has a rather narrow allowed range: $-0.67 <
S_{\pi\pi} < -0.58$. The correlation between $C_{\pi\pi}$ and $S_{\pi\pi}$ is
also to be noted.

2. The unitarity triangle angle $\gamma$ lies in the second quadrant:
$112^\circ < \gamma < 146^\circ$. This range changes a bit if we change the
allowed range of $\sin(2\beta_{eff})$. However, in no case it goes to the
first quadrant, as happens in the pure SM case.  This is understandable: one
needs to have destructive interference between the SM tree and the SM penguin
amplitudes to lower the BR. If one tries to do this entirely with the RPV
amplitude one ends up with unaceptable values of $S_{\pi\pi}$ and
$C_{\pi\pi}$. The above range of $\gamma$ should {\em not} be interpreted as
in conflict with the standard CKM fit results. The reason is that once we have
new physics in $\bbbar$ mixing, the $V_{td}$ and $\sin(2\beta)$ constraints no
longer apply. Also note that the $B\to\pi K$ analysis predicts $\gamma$ in the
second quadrant \cite{bfrs}.

3. The RPV coupling has a new upper bound for 100 GeV sfermions (see
Figure 2),
\begin{equation} 
\label{prod-bnd}
|\lambda'_{i11}\lambda'^*_{i13}| \leq 2.2\times 10^{-3}. 
\end{equation} 
The RPV weak phase lies in the third quadrant if this coupling is
taken to be positive.  This bound is more or less stable against the 
variation of $\beta$. 

4. The strong phase difference between the SM tree and the SM penguin
amplitudes can be kept small, as dictated by different theoretical
models\footnote{Our analysis has been carried out in the context of naive
  factorization \cite{bsw,ali} model. The strong phase is small here, since
  that comes only from the imaginary part of the respective Wilson
  coefficients. It is also small in QCD factorization model. In pQCD
  \cite{keum} it is not so small since the annihilation topologies are taken
  into account.  However, one should note that the pQCD analysis uses $T_c$
  and $P_c$, which are {\em not exactly} identical to our tree and penguin
  amplitudes, respectively.}.  On the other hand, the strong phase difference
between the SM tree and the RPV amplitudes, treated as a phenomenological
parameter, has a very restrictive range: $-40^\circ < \delta < 40^\circ$. The
reason for this restriction is that the CP asymmetries are mainly controlled
by the SM tree and the RPV amplitudes.

\begin{figure}[htbp] 
\vspace{-10pt}
\centerline{\hspace{-3.3mm} 
\rotatebox{-90}{\epsfxsize=8cm\epsfbox{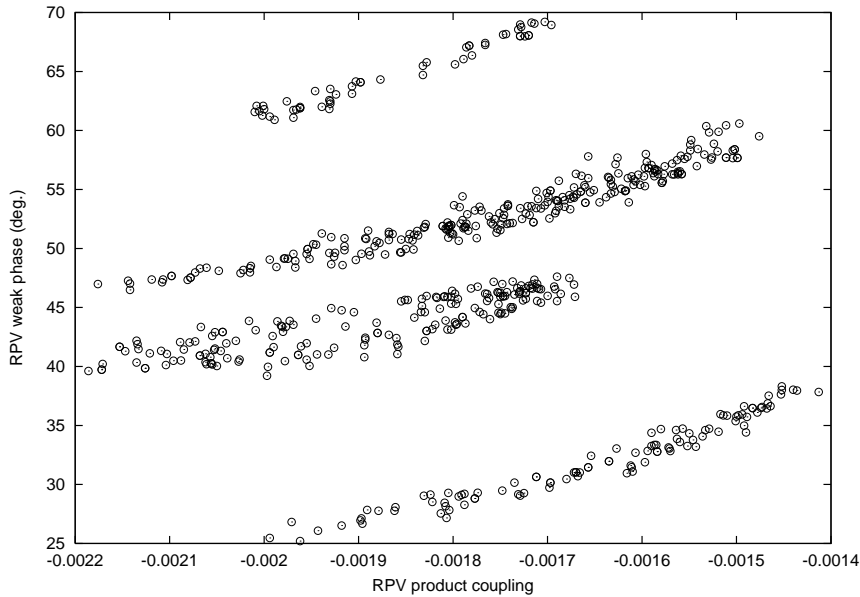}}} 
\hspace{3.3cm}\caption[]{\small Allowed parameter space for the
magnitude and the phase of the product coupling
$\lambda'_{i11}{\lambda'^*_{i13}}$. } \protect\label{fig2}
\end{figure} 

A comment on the robustness of our bound on the product coupling is now in
order. If we vary the experimental data by $\pm 2\sigma$ and allow for the
uncertainties of the involved parameters, the constraint in
Eq.~(\ref{prod-bnd}) relaxes by about 20\%. 
 
So far we have focussed our attention on the quark level process $b\rightarrow
u\bar{u} d$ and studied its impact in $B \r \pi^+ \pi^-$ decay. It is now time
to wonder what would be the impact of the SU(2) conjugate quark level process
$b\rightarrow d\bar{d} d$? Both operators contributes to $B \r \pi^0 \pi^0$
and to understand the nature of the SM and RPV contributions to this process
it is important to recall that the quark composition of $\pi^0$ is the
antisymmetric combination $(u\bar u - d\bar d)/\sqrt{2}$.  In the SM, while
$b\rightarrow u\bar{u} d$ corresponds to a colour suppressed tree diagram,
$b\rightarrow d\bar{d} d$ can proceed only through penguin graphs. It turns
out that a part of the $b\to d\bar{d}d$ RPV amplitude is almost exactly (upto
the SU(2) D-term splitting between sleptons and sneutrinos) cancelled by the
corresponding $b\to u\bar{u}d$ amplitude.  However, another part remains, and
that can significantly enhance the $B\to\pi^0\pi^0$ branching ratio. The
$B^+\to\pi^+\pi^0$ data is satisfied by the allowed RPV parameter space.

The RPV scenario that we have considered in this paper can be directly tested
at colliders. If RPV indeed contributes to $B$ decays as discussed in this
paper, the associated light sleptons/sneutrinos, with masses in the range $100
- 300$ GeV, mediating such decays are very likely to be produced at the
Tevatron and, most certainly, at the LHC. Thus resonant production of
sleptons/sneutrinos \cite{dreiner} will provide a useful cross-check of this
scenario.  The $\lambda'_{i11}$ couplings (in particular, $\lambda'_{211}$)
give rise to a distinct collider signature in the form of like-sign dilepton
signals. Such final states have low SM and $R$-parity conserving supersymmetry
background. The dominant production mechanism is a $\lambda'$ induced resonant
charged slepton production at tree level at hadron colliders. This is followed
by a $R$-parity conserving gauge decay of the charged slepton into a
neutralino and a charged lepton. The neutralino can then decay via the crossed
process to give rise to a second charged lepton, which due to the majorana
nature of the neutralino can have the same charge as the hard lepton produced
in the slepton decay.  It is gratifying to note that the study of
Ref.~\cite{dreiner} shows that for a value of $\lambda'_{211} = 0.05$, which
is perfectly compatible with our bound on the product $\lambda'_{211}
\lambda'^*_{213}$, a smuon mass of about 300 GeV would be visible above the
backgrounds with 2 fb${}^{-1}$ integrated luminosity at the Tevatron Run II,
while for the same coupling a resonant smuon can be observed with a mass of
750 GeV at LHC with 10 fb${}^{-1}$ integrated luminosity.  It is, therefore,
reasonable to expect smuon signals already at the upgraded Tevatron.
 
\section{Conclusions} 
1. RPV contribution to the decay $B\to\pi^+\pi^-$ mode is interesting because
the same new physics amplitude affects both $\bbbar$ mixing and the
decay. Their interplay leads to important constraints. The constraints are
expected to be tighter as more data accumulate.

2. As we have shown, the upper bound on the magnitude of the RPV product
coupling has been improved over the existing results. The RPV weak phase also
gets restricted.

3. A bonus of RPV is that one can enhance the $B\to\pi^0\pi^0$ branching ratio
to a significant level, which is a much desired result in view of what one
finds from the current experiments.

\section*{Acknowledgements}
A.D. and A.K. have been supported by the BRNS grant 2000/37/10/BRNS of
DAE, Govt.\ of India. A.K. has also been supported by the grant
F.10-14/2001 (SR-I) of UGC, India, and by the fellowship of the
Alexander von Humboldt Foundation.  G.B.'s research has been
supported, in part, by the DST, India, project number SP/S2/K-10/2001.

\end{document}